# Comparative Analysis of Deep Learning Architectures for Breast Cancer Diagnosis Using the BreaKHis Dataset

1st İrem Sayın
*Yildiz Technical University*
*Mechatronics Engineering*
Istanbul,Turkey
irem.sayin2@gmail.com

2nd Muhammed Ali Soydaş
*Yildiz Technical University*
*Mechatronics Engineering*
Istanbul,Turkey
ali.soydas@std.yildiz.edu.tr

3rd Yunus Emre Mert
*Yildiz Technical University*
*Mechatronics Engineering*
Istanbul,Turkey
ynsemremert@gmail.com

4th Arda Yarkataş
*SEV American College*

Istanbul,Turkey
ardayarkatas@gmail.com

5th Berk Ergun
*ENKA Schools*

Istanbul, Turkey
berkergun1007@gmail.com

6th Selma Sözen Yeh
*Nort London Collegiate Schoolt*

London, England
selma.sözen.yeh@gmail.com

7th Hüseyin Üvet
*Yildiz Technical University*
*Mechatronics Engineering*
Istanbul,Turkey
huvet@yildiz.edu.tr

*Abstract*— Cancer is an extremely difficult and dangerous health problem because it manifests in so many different ways and affects so many different organs and tissues. The primary goal of this research was to evaluate deep learning models' ability to correctly identify breast cancer cases using the BreakHis dataset. The BreakHis dataset covers a wide range of breast cancer subtypes through its huge collection of histopathological pictures. In this study, we use and compare the performance of five well-known deep learning models for cancer classification: VGG, ResNet, Xception, Inception, and InceptionResNet. The results placed the Xception model at the top, with an F1 score of 0.9 and an accuracy of 89%. At the same time, the Inception and InceptionResNet models both hit accuracy of 87% . However, the F1 score for the Inception model was 87, while that for the InceptionResNet model was 86. These results demonstrate the importance of deep learning methods in making correct breast cancer diagnoses. This highlights the potential to provide improved diagnostic services to patients. The findings of this study not only improve current methods of cancer diagnosis, but also make significant contributions to the creation of new and improved cancer treatment strategies. In a nutshell, the results of this study represent a major advancement in the direction of achieving these vital healthcare goals.

*Keywords— Breast Cancer, Deep Learning, BreaKHis Dataset, Histopathological Images, Transfer Learning*

I. INTRODUCTION (HEADING 1)

One of the major health problems of the 21st century is still cancer. Millions of people throughout the world are afflicted by its many forms, which can present in a wide range of organs and tissues. Among the many forms of cancer, breast cancer has been a major focus of attention, especially among women. The latest numbers from the Global Cancer Observatory are quite discouraging: An astounding 684,996 women lost their lives to breast cancer in 2020[1], out of an estimated 2.3 million new cases diagnosed that year. The need for accurate diagnosis methods and efficient treatment plans is greater than ever.

In the diagnosis of cancer, different diagnostic techniques are used depending on the location and types of cancer. For breast cancer identification, imaging methods such as Mammography, Magnetic Resonance Imaging (MRI), and Computer-Aided Detection (CAD) techniques play a crucial role in the initial steps of diagnosis [2]. CAD systems are specifically designed to assist doctors by highlighting suspicious areas in medical images, thereby enhancing the accuracy and efficiency of diagnosis [3]. In cases where imaging results are suspicious, invasive methods like biopsy are used to obtain definitive results from the suspected cancerous area.

The revolutionary potential of the machine and deep learning approaches in the field of medical diagnostics is getting a lot of momentum in today's high-tech, high-medical-science environment [4-7]. These computational methods, distinguished by their capacity to handle large volumes of data and recognize patterns, provide potential new directions for illness diagnosis and categorization at an early stage [8]. The complexity of breast cancer, with its many subtypes and phases, is confronted head-on by these approaches.

The development of imaging techniques in medicine has greatly aided medical diagnosis. The ability to capture even the smallest of cellular changes in high detail is proving helpful in the battle against illness. The BreakHis dataset is evidence of this development; it is a comprehensive archive of histopathological pictures representing different forms of breast cancer [9]. In addition to providing a detailed look at tumor morphology, this dataset also contains a wealth of information that may be used to train and verify state-of-the-art classification algorithms.

Recent progress in medical diagnostics has been spearheaded by deep learning, a kind of machine learning

[10]. Natural language processing and computer vision are only two of the areas where its designs, such as the popular VGG, ResNet, Xception, Inception, and InceptionResNet, have made waves. These models, with their varied architectural details, provide a fresh perspective from which to examine the complexities of breast cancer cells. Each model's stacked, linked nodes are meant to represent everything from high-level textures and patterns to minute details. By comparing how well various models work on datasets like BreakHis, scientists can learn which ones are most effective in representing the variety of breast cancer presentations.

However, challenges persist in integrating deep learning into medical diagnosis. One primary hurdle is the need for vast amounts of labeled data. There's an evident demand for more diverse and extensive datasets. Additionally, deep learning models require substantial computational resources, making their training time-intensive and demanding on equipment.

Another significant challenge lies in selecting the correct model and fine-tuning hyperparameters. With a plethora of deep learning architectures available, determining the most suitable one for a specific dataset or problem can be daunting. Furthermore, hyperparameter tuning, which involves adjusting various parameters to optimize model performance, can be a complex and time-consuming process. Yet, despite these challenges, the potential improvements in diagnostic accuracy and early detection make the efforts worthwhile

In conclusion, the primary objective of this work is to use the BreakHis dataset to examine the efficacy of deep learning models for breast cancer diagnosis and to identify the best-suited model for this purpose. Histopathological photos of many breast cancer subtypes are included in this dataset. This research intends to categorize different forms of cancer and then assess the efficacy of various deep learning architectures by using models trained on data from publicly available datasets. An integral part of this endeavor is the exploration and determination of the most effective model and hyperparameters for breast cancer diagnosis. The overarching goal of this research is to shed light on how deep learning techniques may be harnessed to enhance the precision of breast cancer diagnoses, ultimately advancing the quality of care and treatment provided to individual patients.

## II. MATERIALS AND METHODOLOGY

### A. Materials

*1) Dataset Description*

The dataset used in this study consists of a collection of microscopic images of breast tumor tissue obtained from a cohort of 82 patients. These images encompass various magnification factors, including 40x, 100x, 200x, and 400x, resulting in a total of 7,909 microscopical images. Known as the BreaKHis database, this dataset was collaboratively established with the P&D Laboratory of Pathological Anatomy and Cytopathology in Parana, Brazil [9]. It encompasses two distinct categories: 2,480 benign and 5,429 malignant tumor samples. Notably, BreaKHis comprehensively covers a spectrum of breast tumor subtypes, classified by expert pathologists into four benign types—adenosis (A), fibroadenoma (F), phyllodes tumor (PT), and tubular adenoma (TA)—and four malignant types—ductal carcinoma (DC), lobular carcinoma (LC), mucinous carcinoma (MC), and papillary carcinoma (PC) (Figure 1).

Each image is represented in the RGB color space, with an 8-bit depth for every channel, and a dimension of 700x460 pixels.

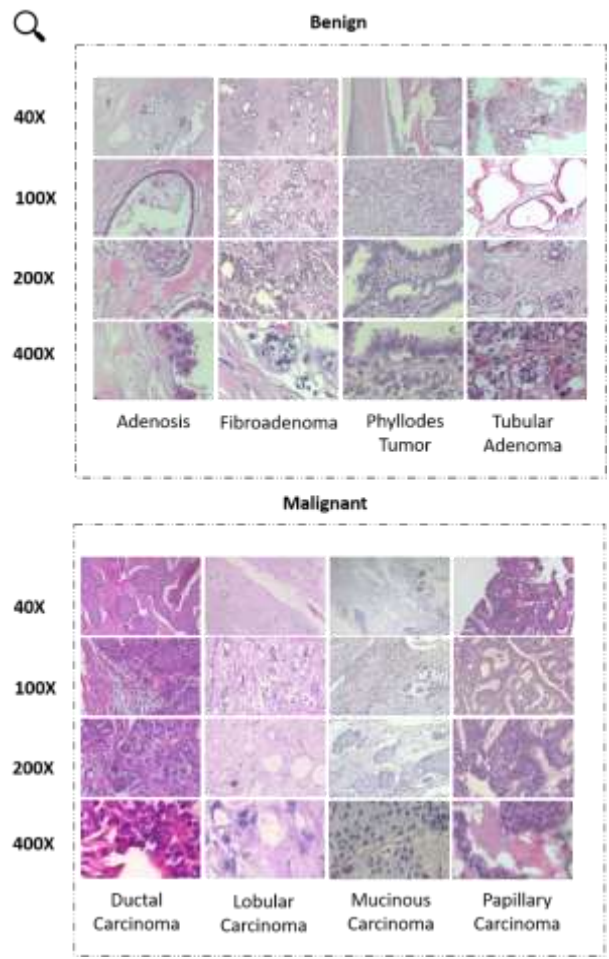

Fig. 1. Sample images from the BreaKHis dataset

*2) Transfer Learning*

Transfer learning, a method that transcends concerns about data volume, is the linchpin of our study. Serving as a pivotal machine learning technique, it improves the creation of models by building on the acquired knowledge of a pre-trained model. This involves optimizing the learning process for a new task using insights from a prior, akin task.

Our approach capitalizes on the efficacy of transfer learning in computer vision tasks. Notably, research demonstrates that knowledge gleaned from extensive image datasets like ImageNet can substantially elevate performance across diverse recognition tasks [11]. This method yields several advantages, including augmented classification accuracy, expedited training, and improved learning performance.

Incorporating transfer learning into our framework necessitates the initial training of a source model on a broad dataset or task. This source model captures overarching features and fundamental attributes. Subsequently, this source model undergoes fine-tuning on the target dataset, which is typically more specific or smaller. During fine-tuning, the model amalgamates its previously acquired general features with task-specific attributes, thereby honing its proficiency for the specific task.

To embark upon our experimental journey, we employed four distinguished deep-learning architectures. Each of these models, including VGG19, ResNet-152, InceptionV3, InceptionResNetV2, and Xception, was initially pre-trained on the expansive ImageNet dataset. This expansive dataset, encompassing over a million labeled images, is instrumental in procuring well-founded pre-trained weights for these models.

*3) Deep Learning Architectures*

Our research harnesses the power of deep learning through the utilization of four prominent Convolutional Neural Network (CNN) architectures: VGG19, ResNet-152, InceptionV3, and Xception. These models have been tailor-made for the intricate tasks of image analysis, feature extraction, and pattern recognition.

The fundamental constituents of these models encompass:

- Convolutional Layer: This layer orchestrates the detection of essential data features. By applying filters to images, it extracts pivotal characteristics. These multi-dimensional filters, composed of pixels, instigate the creation of output matrices, aptly referred to as "feature maps."

- Non-linearity Layer (Activation Layer): Employed after the Convolutional Layer, this layer is recognized for its inclusion of activation functions. These functions, particularly the Rectified Linear Unit (ReLU), enhance the model's non-linear learning capabilities.

- Pooling Layer: This layer is instrumental in diminishing the dimensions of feature maps, accentuating crucial attributes. It significantly reduces computational overhead and parameters by down-sampling and adjusting the stride size.

- Flattening Layer: This layer transforms matrices from Convolutional and Pooling layers into one-dimensional structures, thus preparing the data for the subsequent

- Fully-Connected Layer. Fully-Connected Layer: A cornerstone of deep learning models, this layer is often referred to as the dense layer. It facilitates the establishment of relationships between image features.

*a) VGG19*

The VGG19 model, an abbreviation for Visual Geometry Group 19, has garnered considerable attention in the realms of deep learning and image analysis. Conceived by the "Visual Geometry Group" research team, VGG19 represents a prominent variant within the broader VGG series, comprising a total of 19 layers [12]. This intricate layering architecture plays a pivotal role in enhancing its capacity for perceptive interpretation.

A notable strength of VGG19 lies in its integration of pre-trained weights, which are fine-tuned on the extensive ImageNet dataset to bolster its classification capabilities. During training, VGG19 is fed input data with dimensions of 224x224 pixels, encompassing RGB color channels. This input undergoes a sequential journey through an array of convolutional layers, designed to uncover features using 3x3 pixel filters. This sequence of layers progresses from rudimentary attributes, such as edges and corners, to progressively sophisticated elements, including patterns and more complex structures.

Following the convolutional layers, the model incorporates pooling layers that facilitate a reduction in feature map dimensions while minimizing computational load. Post-pooling, the feature maps are flattened and fed into fully connected layers. These layers serve as conduits for transforming features into higher-level attributes. The final stage often involves the output layer, where probability values are computed to determine the image's classification. These probabilities are then compared with the class labels established during training to evaluate the model's performance [13].

In conclusion, the VGG19 model engages in a meticulous process of feature extraction and classification through its convolutional and fully connected layers. This approach showcases its proficiency, particularly when confronted with voluminous and intricate datasets. The structured architecture, which encapsulates both spatial hierarchies and intricate features, contributes to its capability as a potent tool in image analysis.

*b) ResNet-152v2*

ResNet-152v2, a deep convolutional neural network (CNN) model, holds a significant position within the domain of deep learning, particularly in the realm of image processing. Coined from "Residual Networks," the term "ResNet" is attributed to a pioneering work by Kaiming He and his team in 2015 [14]. This model presents a novel approach aimed at addressing the challenges of elevating the depth of convolutional neural networks (CNNs).

ResNet-152v2, an embodiment of this approach, stands out by leveraging the concept of "residual connections" or "skip connections" to mitigate the issue of vanishing gradients and alleviate the obstacles related to overfitting. These connections enable the propagation of gradients unencumbered by intervening layers, facilitating the training of deeper architectures. The core objective is to empower deeper networks with enhanced learning prowess.

The numeric tag '152' accompanying ResNet signifies its depth, comprising an impressive 152 layers. This voluminous layer count underlines the architecture's profound intricacy, accentuating its potential to encapsulate and decipher complex patterns and features within data. The augmented layer count translates into heightened abstraction and an amplified capacity for feature recognition. However, it is imperative to acknowledge that the pursuit of such depth necessitates larger datasets and computational resources, inevitably posing a trade-off between complexity and resource requisites.

In summation,ResNet-152v2 represents a significant advancement in deep learning, overcoming limitations associated with training deep architectures. This model's propensity to unravel intricate features from data while accommodating increased depth serves as a testament to its commendable contribution in the realm of image analysis and classification.

*c) InceptionV3*

InceptionV3, a deep convolutional neural network (CNN), occupies a pivotal role in the realm of deep learning and image processing. As a member of the Inception family, it was developed by Google Research [15]. This model finds its utility primarily in tasks such as image classification and object detection, where achieving effective outcomes is of paramount importance.

InceptionV3 represents an advanced iteration of the inception architecture, popularly recognized as "GoogleNet." Diverging from conventional convolutional blocks, this architecture adopts "Inception blocks," characterized by multiple convolutions and pooling layers featuring filters of varied dimensions. This innovation facilitates a more efficient capture of patterns and features across different scales.

Manifesting both convolutional and fully connected layers, InceptionV3 is distinguished by its emphasis on a lighter network structure without compromising performance. This pursuit aims to attain superior results with reduced parameters and computational complexity compared to broader and deeper network architectures.

InceptionV3 frequently finds application in scenarios involving extensive datasets and intricate visual tasks. Notably, it excels in domains such as object recognition, classification, segmentation, and image style transfer, showcasing its proficiency in producing impactful outcomes.

In conclusion, InceptionV3 stands as a testament to the iterative refinement of deep neural networks, advancing the capabilities of image analysis and interpretation. Its utilization within the landscape of sizable data and multifaceted visual tasks underscores its efficacy, making it a potent asset in the arsenal of modern deep-learning methodologies.

*d) Xception*

Xception, an advanced convolutional neural network (CNN), stands as a notable contender in the realms of deep learning and image processing. Also recognized as "eXtreme Inception," Xception is an innovation attributed to Google [16]. It represents a refined adaptation of CNNs, strategically aimed at superior feature extraction from image data. This architecture outperforms its predecessor, Inception V3, particularly on the ImageNet dataset, comprising 350 million images spanning 17,000 categories.

Central to the Xception architecture are the "Depthwise Separable Convolution" layers, characterized by a two-fold procedure:

- Depthwise Convolution: Employing compact convolutional kernels, this stage independently processes input channels, significantly curtailing computational complexity and learnable parameter count.
- Pointwise Convolution: Subsequently, 1x1 convolutional kernels facilitate interactions among channels, fostering dynamic relationships and intricate feature combinations.

Based on the Inception architecture, Xception advances the convolutional paradigm. It enhances Inception's multidimensional convolutional and pooling layers, or "Inception blocks," aiming for finer-grained convolutional operations and more effective feature extraction.

Xception's convolutional operations unfold in two stages: "cross-channel correlation" and "spatial correlation," designed to capture inter-channel relationships and yield comprehensive and intricate features.

Prominent in image-centric convolutional neural networks (CNNs), Xception excels in handling extensive datasets and intricate image characteristics. Its efficacy shines in image classification, object detection, and segmentation, among other visual tasks.

As an evolved iteration of the Inception architecture, Xception occupies a significant role in image classification, object detection, and segmentation. Its distinct attributes contribute substantially to the enhancement of convolution-based models, signifying a pivotal step towards elevating performance benchmarks.

*e) Inception-ResNetV2*

Inception-ResNetV2 emerges as a pivotal convolutional neural network (CNN) model profoundly utilized within the realms of deep learning and image processing. Conceived as a collaborative effort by Google, this model synergistically amalgamates the innovative features of the Inception and ResNet architectures [17]. By amalgamating the salient concepts from both Inception and ResNet paradigms, Inception-ResNetV2 aims to achieve heightened performance and feature extraction capabilities.

In essence, Inception-ResNetV2 can be perceived as a hybrid offspring of the InceptionV4 and ResNet architectures. Rooted in the foundation of the Inception paradigm, the architecture features intricately composed "Inception blocks." These blocks incorporate a diverse ensemble of convolutional layers, endowed with varying filter sizes and pooling operations. This orchestration empowers the network to imbibe an array of dimensions, thus facilitating the acquisition of diverse and multi-scale features.

The ResNet architecture, in contrast, draws its efficacy from the ingenious integration of residual connections. This revolutionary concept mitigates the hurdles associated with vanishing gradients and facilitates more efficacious training of deep networks. By harmonizing the outputs of a given layer with those of a preceding layer, the model benefits from optimized learning dynamics.

Incorporating the quintessence of InceptionV4's attributes and fusing them with the depth-driven learning capabilities and regularization proficiency of the ResNet paradigm, Inception-ResNetV2 endeavors to culminate in an enhanced performance paradigm. With a deeper architectural disposition, the model endeavors to derive superior performance metrics, aligning with the demands of intricate data representations.

In practice, Inception-ResNetV2 finds its forte in the domain of image processing, particularly well-suited for the rigors of expansive and intricate datasets. Its application spectrum encompasses a plethora of tasks, including but not limited to object recognition, image classification, segmentation, and the innovative pursuit of image style transfer.

In summary, Inception-ResNetV2 stands as an exemplar of the dynamic interplay between architectural paradigms in deep learning. By unifying the virtues of Inception and ResNet, this model encapsulates a rich arsenal of capabilities, well-equipped to address the challenges posed by complex and varied image processing tasks.

In our research, we harnessed the prowess of these architectures to tackle the intricate task of breast tumor classification using the BreaKHis dataset. By leveraging transfer learning, we adapted these pre-trained models to categorize breast tumor images based on their histopathological types. The ensuing section delineates the outcomes of our experiments and the subsequent results of these used models.

### B. Methodology

#### 1) Training Environment

Deep learning models were trained on an 11th generation Intel(R) Core(TM) i7-11700KF processor coupled with an NVIDIA GeForce RTX 3090 GPU.

#### 2) Dataset Description and Augmentation

The primary dataset used for this research was composed of images magnified 200 times. Out of the entire collection of 1,994 images, they were strategically divided into three distinct sets: 1,270 images (or 65% of the total) were used for training, 255 images (15%) for validation, and the remaining 499 images (20%) were reserved for testing. To enhance the robustness of the models and improve their generalization capabilities, data augmentation techniques were applied. Every image, except those in the test set, underwent several transformations: they were expanded and elongated by a factor of 0.2, subjected to a 30-degree rotation, and also flipped along the horizontal axis.

#### 3) Model Architectures and Training

Several renowned deep learning architectures were employed in this study, including VGG16, VGG19, ResNet152V2, InceptionV3, InceptionResNetV2, and Xception. Each of these models was fine-tuned to better adapt to our specific dataset. It's worth noting that our dataset was unique and didn't rely on any pre-established industry database. This uniqueness posed challenges, especially since traditional sampling methods were insufficient. To address this, a systematic approach was adopted where layers of each model were made trainable in a specific sequence. After each training step, the model's performance was evaluated, and the confusion matrix was recorded. This iterative process helped identify the optimal configuration for each model based on the best performance metrics.

#### 4) Techniques Applied Across Models

Three pivotal techniques were consistently applied across all models to ensure optimal training:

- Early Stopping: Given the limited size of our dataset, there was a risk of models overfitting or learning the training data too closely. To mitigate this, an 'early stopping' technique was employed. This technique monitors the model's performance on the validation set and halts training once the model starts to overfit, ensuring that the models generalize well to new, unseen data. The threshold for this was set at a minimum learning rate of 1e-7.

- Learning Rate Adjustment: During training, there are instances when the model's learning progress slows down or plateaus. To counteract this and rejuvenate the learning process, the learning rate was dynamically adjusted using the ReduceLROnPlateau function. The floor for this adjustment was also set at a learning rate of 1e-7.

- Class Weight Balancing: An imbalance was observed in the dataset, where images of the 'DC' class were significantly more numerous than others. This could bias the model towards the 'DC' class. To ensure all classes were treated fairly during training, a 'class weight' method was applied, setting the mode to 'balanced'.

#### 5) Detailed Model Configurations:

##### a) VGG16 and VGG19

Both of these models utilized a Sequential structure. After importing the pre-trained base models, additional layers were appended to adapt to our dataset. These included a Flatten layer to reshape the data, followed by a Dense layer for complex pattern recognition, and a Dropout layer to prevent overfitting. The models were then capped with an output layer equipped with a Softmax activation function for classification. For both models, the last 15 layers were made trainable. They were trained for up to 50 epochs, using batches of 32 images, and an input resolution of 224x224 pixels. With the early stopping mechanism, the VGG16 model's training concluded at 33 epochs, while the VGG19 model finished at 25 epochs.

##### b) ResNet152V2

This model was built upon the pre-trained ResNet152V2 base. After obtaining the base model's output, a GlobalAveragePooling2D layer was added to simplify the data structure, followed by a Dense layer and a Dropout layer for pattern recognition and regularization, respectively. The model's architecture was completed with an output layer using a Softmax activation for classification. The last 200 layers of this model were set to trainable. The training regimen consisted of 30 epochs, batches of 32 images, and an input resolution of 224x224 pixels. The early stopping mechanism concluded the training at 21 epochs.

##### c) InceptionV3, InceptionResNetV2 and Xception

For InceptionV3, InceptionResNetV2, and Xception, pre-trained base models served as the foundation. The output from these base models was processed using a GlobalAveragePooling2D layer. This was followed by a Dense layer for pattern recognition. Additionally, a Dropout layer was added for regularization (except for InceptionV3). Each model was finalized with an output layer equipped with a Softmax activation for classification. The models had varying numbers of trainable layers: 260 for InceptionV3, 720 for InceptionResNetV2, and 115 for Xception. All three models were trained for 30 epochs, with batches of 32 images. However, they used a slightly larger input resolution of 299x299 pixels. The early stopping mechanism concluded training at 23 epochs for InceptionV3, 17 epochs for both InceptionResNetV2 and Xception.

## III. RESULTS

To accurately assess the performance of the deep learning architectures employed in this study, several evaluation metrics were utilized. A brief description of these metrics and their respective calculations is provided below:

- Accuracy: Defined as the ratio of correctly predicted instances to the total instances in the dataset.

- Precision: Precision quantifies the ratio of correctly predicted positive observations to the total predicted positives.

- Recall (Sensitivity): Recall measures the ratio of correctly predicted positive observations to all the observations in the actual class.

- F1 Score: Representing the harmonic mean of Precision and Recall, the F1 Score offers a balance between the two metrics.

For a detailed breakdown of the equations and definitions of these metrics, please refer to Table 1.

TABLE I. EVALUATION METRICS

| Evaluation metrics | Formula |
|---|---|
| Accuracy | $Accuracy = \frac{Number\ of\ Correct\ Predictions}{Total\ Number\ of\ Predictions}$ |
| Precision | $Precision = \frac{True\ Positives}{True\ positives + False\ Positives}$ |
| Recall | $Recall = \frac{True\ Positives}{True\ Positives + False\ Negatives}$ |
| F1 Score | $F1\ Score = 2 \times \frac{Precision \times Recall}{Precision + Recall}$ |

With a clear understanding of these metrics, the performance of the deep learning architectures on a test set of 499 images is detailed as follows:

The VGG16 model demonstrated an accuracy of 65%, complemented by a precision of 61% and a recall of 68%. Its macro F1 score stood at 63%, with the weighted F1 score slightly higher at 65%. Moving to the VGG19 model, it achieved an accuracy of 69%, a precision of 64%, and a recall of 58%. The corresponding macro and weighted F1 scores were 58% and 67%, respectively.

The ResNet152V2 model, in contrast, showcased a robust performance with an 80% accuracy, 79% precision, and an 82% recall. Both its macro and weighted F1 scores were consistent at 80%. The InceptionV3 model further elevated the benchmarks, registering an accuracy of 87%, precision of 85%, and a recall of 89%. Its macro and weighted F1 scores were both impressive at 87%. Similarly, the InceptionResNetV2 model achieved an accuracy of 87%, but with a slightly higher precision of 90% and a recall of 85%. Its macro F1 score was 86%, and the weighted F1 score was 87%.

Concluding with the Xception model, it emerged as a top performer, boasting an accuracy, precision, and recall all at 90%. Both its macro and weighted F1 scores were recorded at 89%.

Comparatively, the Xception model stands out as the best performer across most metrics, while the VGG16 and VGG19 models, despite their contributions, lagged in performance. This comparative analysis underscores the Xception model's potential for breast tumor classification tasks, but it's essential to align model selection with specific application requirements.

In a comparative analysis, as summarized in the Table 1 below, the Xception model emerged as the best performer in terms of accuracy, precision, recall, and F1 scores. On the other hand, while the VGG16 and VGG19 models provided valuable insights, they lagged behind the other models in most metrics.

TABLE II. COMPARISON OF THE MODELS

| Model | Accuracy | Precision | Recall | Macro F1 | Weighted F1 |
|---|---|---|---|---|---|
| VGG16 | 0.65 | 0.61 | 0.68 | 0.63 | 0.65 |
| VGG19 | 0.69 | 0.64 | 0.58 | 0.58 | 0.67 |
| ResNet152V2 | 0.8 | 0.79 | 0.82 | 0.8 | 0.8 |
| InceptionV3 | 0.87 | 0.85 | 0.89 | 0.87 | 0.87 |
| InceptionResNetV2 | 0.87 | 0.9 | 0.85 | 0.86 | 0.87 |
| Xception | 0.89 | 0.9 | 0.9 | 0.9 | 0.89 |

In conclusion, while each model has its strengths and nuances, the Xception model's performance underscores its potential for breast tumor classification tasks. However, it's essential to consider the specific requirements and constraints of any application before selecting an appropriate model.

IV. DISCUSSION

Insightful knowledge of the strengths and weaknesses of several deep learning architectures for breast cancer diagnosis has been provided by this work. A novel insight into these models' capabilities in a real-world medical setting is provided by the incorporation of transfer learning and the use of the BreaKHis dataset.

It's notable how well the Xception model performs compared to others. Its design, which is a refinement of the Inception model, appears to be well-suited to the complexities of histopathology pictures. The depthwise separable convolutions at the heart of Xception's architecture have the potential to play a pivotal role in the algorithm's success. By employing these convolutions, the model is better able to capture spatial and cross-channel correlations, which are essential for identifying nuances across breast cancer subtypes.

While Xception did the best of the models tested, it is important not to discount the work done by the competition. Examples of models with impressive performance are InceptionV3 and InceptionResNetV2. In the instance of InceptionResNetV2, the residual connections complement the Inception blocks to create a well-rounded method for feature extraction and pattern detection.

The simplified architecture of VGG16 and VGG19 may explain why they perform less well than the other models. Despite their usefulness in many image processing applications, histopathological pictures may benefit from more sophisticated structures due to their complexity and diversity. More than that, these models are more prone to overfitting because to the vast number of trainable parameters, especially in the absence of a suitably big and varied training dataset.

It's important to remember the obstacles that had to be overcome during this research. Deep learning presents considerable challenges in the field of medical diagnostics due to the high processing requirements, the risk of overfitting, and the necessity for high volumes of labeled data. The application of techniques like early stopping, learning rate adjustment, the addition of dropout layers and class weight balancing were crucial in navigating these challenges.

## V. CONCLUSION

Breast cancer remains a significant health challenge, and the quest for accurate and early diagnosis is paramount. This study underscores the transformative potential of deep learning methodologies in breast cancer diagnosis. Among the models evaluated, the Xception architecture emerged as a particularly promising tool, demonstrating superior performance in classifying breast tumor subtypes.

However, the journey of integrating deep learning into medical diagnostics is an ongoing one. As technology evolves and more data becomes available, there's potential for even more accurate and efficient models. Future research might delve deeper into hybrid models, combining the strengths of multiple architectures, or explore entirely new methodologies. Additionally, while this study focused on the 200x zoom data from the BreakHis dataset, future investigations could benefit from utilizing other magnification levels and experimenting with different augmentation methods to enhance model robustness and generalizability.

It's also crucial to remember that while these models can aid diagnosis, the expertise and judgment of medical professionals remain irreplaceable. Deep learning models should be viewed as tools that complement, rather than replace, the skills of pathologists and oncologists.

In the broader perspective, the fusion of technology and medicine holds the promise of revolutionizing patient care. With continued research and collaboration between the tech and medical communities, there's hope for a future where diseases like breast cancer can be diagnosed and treated with unprecedented precision.